\documentclass{aastex}
\usepackage{spr-astr-addons}
\usepackage{url}
\usepackage{epsfig}
\usepackage{ulem}

\RequirePackage{color}
\definecolor{blue(colorwheel)}{rgb}{0.0, 0.5, 1.0}

\begin{document}

\title{Comets in UV}
\slugcomment{Not to appear in Nonlearned J., 45.}
\shorttitle{Short article title}
\shortauthors{Autors et al.}

\author{B. Shustov\altaffilmark{1}} \and \author{M.Sachkov\altaffilmark{1}}
\affil{Institute of Astronomy of the RAS, Russia}
\and
\author{Ana I. G\'omez de Castro \altaffilmark{2}} \and \author{Juan C. Vallejo\altaffilmark{2}}
\affil{AEGORA Research Group, Universidad Complutense, Spain}
\and
\author{E.Kanev\altaffilmark{1}}
\affil{Institute of Astronomy of the RAS, Russia}
\and
\author{V.Dorofeeva\altaffilmark{3,1}}
\affil{Institute of Geochemistry of the RAS, Russia}

\altaffiltext{1}{Institute of  Astronomy, Russian Academy of Sciences, 
	Pyatnitskaya 48,  119017 Moscow, Russia}
\altaffiltext{2}{AEGORA Research Group, Fac. CC. Matematicas, Universidad Complutense, Plaza de Ciencias 3, 28040 Madrid, Spain}
\altaffiltext{3}{Vernadsky Institute of Geochemistry and Analytical Chemistry, Russian Academy of Sciences, Kosygina 19, 119991 Moscow,  Russia}

\begin{abstract}
Comets are  important ``eyewitnesses'' of   Solar System formation and evolution. Important tests to determine the chemical composition and to study the physical processes in cometary nuclei and coma need data in the UV range of the electromagnetic spectrum. Comprehensive and complete studies require  for additional ground-based observations and in-situ experiments.  We briefly review observations of comets in the ultraviolet (UV)  and  discuss the prospects of UV observations of comets and exocomets  with space-born instruments. A special refer is made to the World Space Observatory-Ultraviolet (WSO-UV) project.   
\end{abstract}



\section{Introduction}

Comets are common and very diverse minor bodies of the Solar System. They are common in extrasolar planetary systems too. Comets are very interesting objects in themselves. A wide variety of physical and chemical processes taking place in cometary coma make of them excellent space laboratories that help us understanding many phenomena not only in space but also on the Earth.

We can learn about the origin and early stages of the evolution of the Solar System analogues, by watching circumstellar protoplanetary disks and planets around other stars. As to the Solar System itself comets are considered to be the major ``witnesses'' of its formation and early evolution. The chemical composition of cometary cores is believed to basically represent the composition of the protoplanetary cloud from which the Solar System was formed   approximately 4.5 billion years ago, i.e. over all this time the chemical composition of cores of comets (at least of the long period ones) has not undergone any significant changes. 

At present, in the outer part of the Solar System two populations of small bodies are  considered to be sources of the observed comets: the  trans-Neptunian region, consisting of the Kuiper belt and the scattered disc (30 -- 100\,AU) and the Oort cloud (up to 100\,000\,AU ). In recent years, the Main asteroid belt is also considered  as a source for main-belt comets. In many models the origin of these populations is associated with processes that accompany the formation of the planetary system (see e.g. \cite{Dones2015, Meierhenrich2015}).

We believe that careful analysis of the chemical composition of comets  with  adequate chemical model of early Solar System (protoplanetary disk)  could help to restore information about  area of the disk where these bodies have been formed. By linking this information with dynamic models of the evolution of Solar System, we can verify our understanding of the early Solar System.

\begin{figure*}
	\centering
	\includegraphics[width=0.8\linewidth]{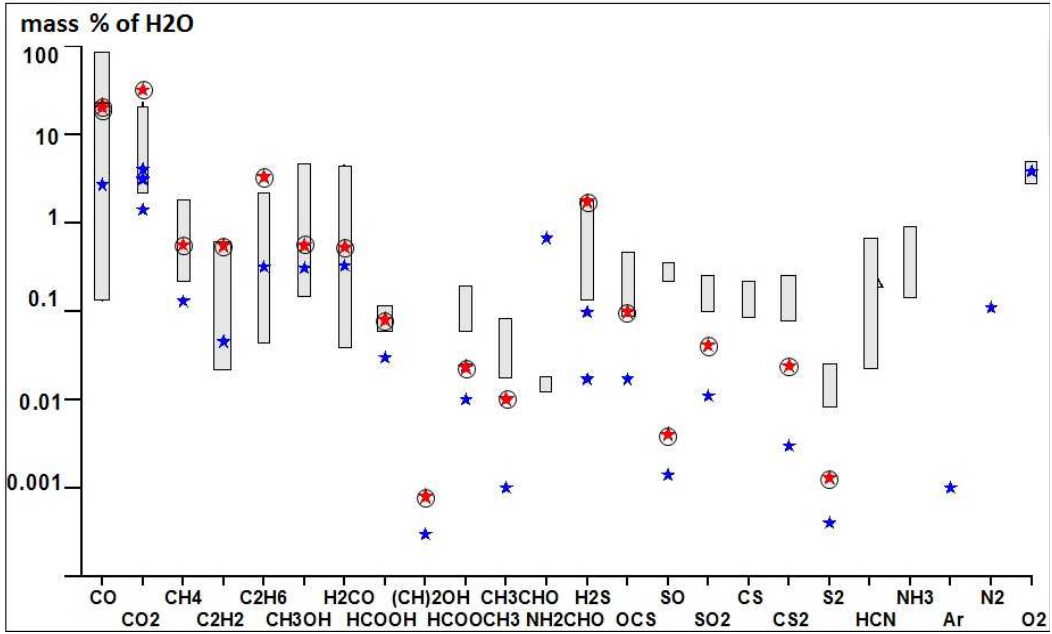}
	\caption{Values of relative contents of gas components in the coma of long-period and short-period comets by percent of water mass. For 67P/Churyumov-Gerasimenko  comet the gas contents are shown by black asterisks (before perihelion) and by encircled asterisks after
perihelium passage. See text for references.}
	\label{fig1}
\end{figure*}

At present, data on the comet composition of about twenty short-period and long-period comets can be found in the literature (\cite{Bockelee-Morvan2011, Mumma2011, Ootsubo2012, Paganini2014, Cochran2015, Roth2017}). This data are presented in Fig.\ref{fig1}. Analysis of this data shows that there are no fundamental differences between the volatiles content in comets of different dynamic types. 
Of course,  we should bear in mind that only the very deep interiors ($>50$\,m) of cometary nuclei would have information on the primary composition of cometesimals. 

Observations of cometary interiors are challenging.
Since we are  limited by the distant observations, a critical problem is distinguishing between the coma and the nucleus signatures. Chemical composition of the matter evaporated from the surface of the comet (i.e. coma) is typicaly far from  primordial. The upper layers of the core of short period comets are intensively reproccessed during previous encounters with the Sun. 
This is confirmed by the data obtained for  comet 67P/C-G in the Rosetta experiment (\cite{LeRoy2015,Rubin2015,Balsiger2016,Altwegg2016,Mall2016}). This data are also represented in Fig. 1: asterisks (before perihelion) and by encircled asterisks (after perihelium).  All the volatile components are stronger after perihelion passage, when the upper shielding layer is  destroyed in the southern hemisphere of the comet. From the analysis of the data and considering the propagation velocity of heat within the nucleus, it can be concluded that observational data obtained 2 -- 3 weeks after the perihelion comet is of particular value, since they will most adequately correspond to the composition of the cometary nucleus.

Many problems of determining the chemical composition of comets  can only be solved by using observational data in the ultraviolet (UV) range of the electromagnetic spectrum. Cometary spectra in the UV and visible range include lines  formed by scattering  of the sunlight on solid dust particles and emission lines from the gas in coma. UV spectroscopy provides important information about the composition of a cometary coma. In general, cometary coma are dominated by emission features that originate from the dissociation products of water: OH, H and O, or correspond to secondary atomic, molecular, and ionic species, such as C, C$^+$, CO, CO$^+$, CO$_2^+$, S, and CS. Also, the spectra and images contain information on structure and dynamical features of coma.  The spatial structure   and time variations of the outflows inferred from the images and spectra help to reveal the structure and other properties of cores. There are ways to restore the thermal history of cores too. For instance, N$_2$ and  noble gases are both chemically inert and highly volatile (evaporate at very low temperatures), so they are particularly interesting  clues on the comets thermal history. But they have not yet been detected by any UV instrument in cometary comas.

 \cite{Delsemme1980} reminded that the atmospheric bands of ozone absorb strongly at
wavelengths less than 300\,nm and still absorb irregularly
up to wavelengths longer than 320\,nm. 
 The 300 -- 320\,nm range (where the famous OH line at $\lambda$309\,nm)  lies  is a very important ``interface domain'' typically covered by space-born UV instruments and  still reachable by groud-based telescopes located in mountains. This makes possible various tests and calibrations.

In this paper we discuss findings  from UV (far UV (FUV) and near  UV  (NUV)) observations of comets with space born instruments (Section 2).   Special attention is paid to the prospects of the World Space Observatory -- Ultraviolet (WSO-UV) project  (Section 3).  Concluding remarks are given in Section 4.

\section{Half a century of UV-observations of comets from space}

In an interesting  review by \cite{Delsemme1980} of early UV observations of comets,  all spectral {\it first-time} identifications of comets are classified  in three periods: 

\begin{itemize}
	\item From the early times until 1911 (named by \cite{Delsemme1980} as Dark Ages of Cometary Spectra), when the two major features of the neutral coma spectrum, CN and C2, had been already identified as well as the two major ions of
	the ion tail (CO$^+$ and N$_2^+$). Metallic lines had already
	been seen in a sun-grazing comet, and the first UV features below 340 nm had just been observed as a fuzzy unresolved band.  Next thirty years was a period with no bright achievements in the field,

	\item Period 1941 -- 1970 started with the first definite detection by \cite{Swings1941} of OH(0-0) at 309\,nm and NH(0-0) near 336\,nm in the spectrum of  comet Cunningham (1941I). The  discovery of the UV line of OH   demonstrated that comets contain water. This was strongly confirmed by the	first rocket and satellite experiments running UV observations of comets. The  ``space era'' of UV studies of comets began about half a century ago. 
	
	\item  Since then, we are in the  ``modern'' epoch in UV studies. A number of  space-born instruments (few to 240\,cm aperture, both spectroscopic and imaging ones) were (and are) used for numerous UV observations of comets.  We think this period will continue  up to the appearance of giant UV telescopes with aperture of 4 -- 30\,m. This future epoch will come not earlier than  20 years from now and currently, the most powerful tools are telescopes of  the 2\,m class (HST, WSO-UV).
	
\end{itemize}

In the article by \cite{Delsemme1980}, the first two periods are described in some detail.   Here we briefly discuss the most important scientific and technological achievemets  of the third  epoch as well as some prospects for the coming decade. We think that the major directions for the UV observation of comets mentioned by \cite{Delsemme1980}   remain important in our days. These are: observations of the Lyman-$\alpha$ (Ly$\alpha$) halos of comets, OH observations   and observations of C, 0, S atoms, ions, and molecules. Both imaging and spectroscopic technologies are used.

\subsection{Imagery and low resolution spectroscopy}

Imagery and low resolution spectroscopy were historically the most popular technologies at the  dawn of UV ``space era''.

\subsubsection {Rocket and balloon UV experiments}

First rocket and satellite experiments on UV observation of comets immediately brought the very impressive  discovery of the immense (up to 1.5 million\,km across) Ly$\alpha$ halo of comets: Tago-Sato-Kosaka 1969g (C/1969\,T1) by the OAO-2 satellite (\cite{Code1970})  and an Aerobee sounding rocket (\cite{Jenkins1972}); comet Bennett (1969i) (\cite{Bertaux1970}), comet Encke by the OAO-2 satellite and the OGO-5 satellite (\cite{Bertaux1973}). 
Comet West (1976\,VI) was observed with  an Aerobee rocket payload (instruments):  spectra   (\cite{Feldman1976}) and imagery of the Ly$\alpha$ halo  (\cite{Opal1977}) were obtained. These are some of the first observations of comets in  UV.

Rocket experiments remain important in our days.   
FUV imagery and objective grating spectroscopy of comet C/2012 S1 (ISON) were acquired from NASA sounding rocket 36.296 UG launched in 2013 (\cite{McCandliss2015}). The comet was $0.1^{\circ}$ below ground horizon, 0.44\,AU from the Sun. The payload reached an apogee of 279\,km and the total time pointed at the comet was 353\,s.  A wide-field multi-object spectro-telescope called FORTIS (Far-UV Off Rowland-circle Telescope for Imaging and Spectroscopy) observed the Ly$\alpha$ emission making use of  an objective grating mode through an open microshutter array over a 0.5 square degree  field of view. Hydrogen production rate was estimated to be $\sim5\times10^{29}$ atoms s$^{-1}$. They also acquired a broadband image of the comet in the 128 -- 190\,nm bandpass.

An obvious disadvantage of rocket experiments is the short exposure time. This is  overcome by balloon missions. The UV window has been scarcely explored  through balloons.  An  UV-VIS (Ultra-Violet/Visible) instrument was designed for the BOPPS mission (Balloon Observation Platform for Planetary Science) \cite{Young2014}.  BOPPS was launched in 2014 and observed comets, using a 0.8\,m aperture telescope, a pointing system that achieved better than  1'  pointing stability, and an imaging instrument suite covering from the near-ultraviolet to the mid-infrared. BOPPS observed two Oort cloud comets: C/2013\,A1 (Siding Spring) and C/2014\,E2 (Jacques). The major results were received with the IR instruments (\cite{Cheng2017}).

\subsubsection {UV observations onboard interplanetary missions}
Many planeatry missions have been equipped with UV instruments that were capable to observe comets.  We mention few of them. 

The Ly$\alpha$ halo of comet Kohoutek was repeatedly
observed by the UV spectrometer of the
Mariner\,10 spacecraft during its transit to planet
Venus (\cite{Kumar1979}) and by different instruments on board Skylab,
namely, the electrographic Schmidt camera (110--150 
nm)  (\cite{Carruthers1974}), the high resolution spectrograph  (\cite{Keller1975}). 

\cite{Crismani2015} reported that 
NASA's Imaging Ultraviolet Spectrograph (IUVS) ( $\lambda\lambda$110 -- 340 nm, resolution 0.5\,nm) onboard the Mars Atmosphere and Volatile EvolutioN (MAVEN) orbiting spacecraft obtained images of the hydrogen coma of the planet-grazing comet C/2013 A1 (Siding Spring) 2 days before its close encounter with Mars. Observations by IUVS brought interesting results: the water production rate was  $1.1\pm0.5\times10^{28}$ molecules/s, the total impacting fluence of atoms and molecules corresponding to the photodissociation of water and its daughter species was $2.4\pm1.2\times10^{4}$kg. These observations were used to confirm predictions that the mass of delivered hydrogen is comparable to the existing reservoir above 150\,km. 

Spectral observations of Comets C/2012\,S1 (ISON) and 2P/Encke were acquired by the MESSENGER spacecraft during the close passes of both comets by Mercury (\cite{Vervack2014}).

Several campaigns of cometary observations have been performed by the UV spectrometer SPICAV from 2012 to 2014. SPICAV is one of the Venera-Express instruments.
The UV channel of SPICAV is a full UV imaging 40\,mm aperture spectrometer for  
 118 -- 320\,nm,  $\Delta\lambda \sim 0.5$\,nm. 
For the 6 observed comets  water sublimation rates as a function of the Sun distance were estimated (\cite{Chaufray2015}).

Comet 2P/Encke was observed with UVCS/SOHO near perihelion (2000 September 9 and 11) in the Lyman lines of hydrogen. \cite{Raymond2002}  
reported that the narrow Ly$\alpha$  profile indicates that the observed photons are scattered from hydrogen atoms produced by dissociation of H2O and OH, though a broader profile far from the coma suggests a contribution from hydrogen atoms produced by charge transfer with solar wind protons.

The relative paucity of solar UV photons has limited the observations from orbiting observatories. 
The highest spatial resolution  has been obtained  with the Space Telescope Imaging Spectrograph (STIS) on HST, but is typically limited to tens of
kilometers for the comets passing the closest to the Earth.
The ALICE FUV spectrograph onboard the Rosetta mission made possible
observations on scales of tens to hundreds of meters near
the nucleus of the comet. ALICE is an  imaging spectrograph in the $\lambda\lambda$70 -- 205\,nm range, with  resolution 0.8 -- 1.2\,nm. The slit is $5.5^{\circ}$ long with a width of 0.05$^{\circ}$ --  0.10$^{\circ}$. ALICE is  an off-axis telescope feeding a 0.15-m normal incidence Rowland circle spectrograph with a concave holographic reflection grating. The MCP detector used solar-blind opaque photocathodes (KBr and CsI).
 \cite{Feldman2015}  reported the results of the observations beginning in August 2014, when the spacecraft entered an orbit  about 100\,km above the nucleus. These observations were an unique opportunity to probe the coma -- nucleus
interaction region that is not accessible to the in situ instruments on Rosetta or to observations from Earth orbit. 

\subsubsection {UV observations of comets onboard general purpose (astrophysical) missions}

\begin{figure*}[ht!]
	\centering
	\includegraphics[width=0.8\linewidth]{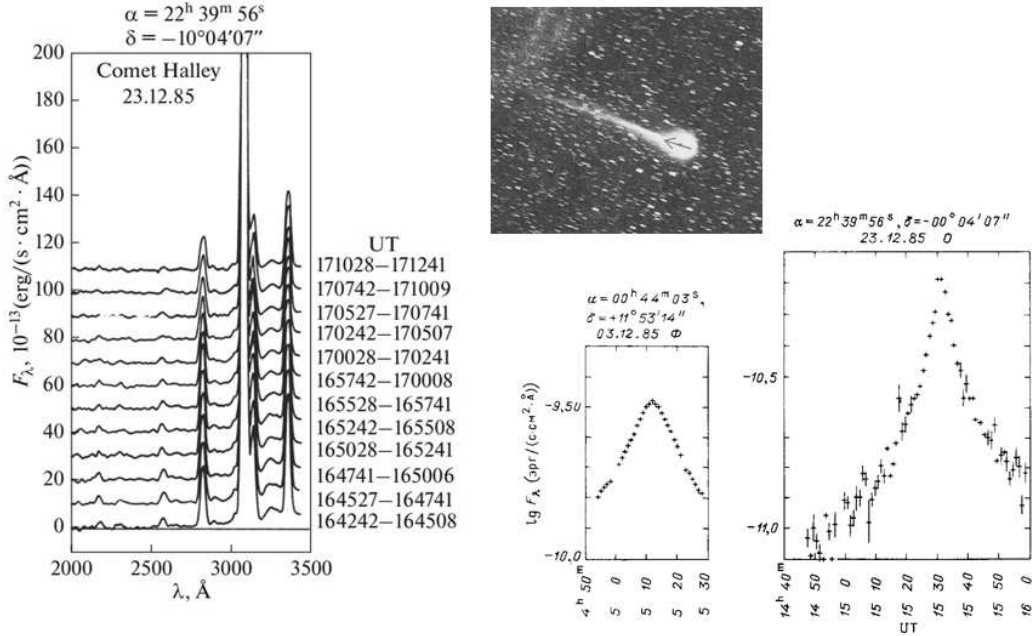}
	\caption{Fragment of comet Halley UV spectra obtained by the   mission "Astron'' in 1986 (\cite{Boyarchuk1994}). Photometic cuts in OH 
made in direction shown by arroware shown either.}
	\label{fig2}
\end{figure*}

Observations of the Ly$\alpha$ halos of comets  continued in forhcoming experiments. We briefly describe some examples.

The Copernicus satellite
spectrometer scanned the Ly$\alpha$ halo of comets
Kohoutek and Kobayashi-Berger-Milon (\cite{Festou1979}).

 \cite{ Kaneda1986}) presented results of 
Ly$\alpha$ imaging of  Ly$\alpha$ Halley comet  performed with S/C Suisei.   
The strong ``breathing'' of comet Halley coma with a period of 2.2 days was monitored throughout all phases in the 1985 -- 1986  perihelium passage. Concerning to the water production, postperihelion rates exceeded the preperihelion ones by factor of  2 or 3. This fact suggests a postperihelion heating of the cometary nucleus. Some fine structures in the Ly$\alpha$ photometry data taken at the encounter indicate spatial variations of the atomic hydrogen density in the coma. These may correspond to the shell structures in the images so far obtained and be ascribed to the presence of a hydrogen source other than H$_2$O in the cometary gas or dust particles. This second hydrogen source is expected to have a larger photo-dissociation rate than H$_2$0.

 Molecules of OH were  repeatedly observed  in  comets. For instance,  the OH bands in spectra of on comets Bennett, Kohoutek and  periodic comet Encke were  observed with the NASA Convair 990 aircraft
especially equipped for this purpose. The OH bands are very strong in all UV spectrograms of comets that covered the proper range (\cite{Harvey1974}).

Comet Halley was observed by the Soviet  UV observatory  ``Astron'' with the  80\,cm telescope ``Spika'' onboard (the largest space telescope at that time). The telescope was  equipped with a French scanning spectrometer (\cite{Boyarchuk1986,Boyarchuk1994}). The very  strong lines of  OH  at 309 nm and NH at 336 nm seen in the spectrograms are shown in Fig.\ref{fig2}. 

Comet Seargent (1978\,m), the first comet to be observed
by the International Ultraviolet Explorer (IUE) satellite,
resolved the OH (0-0) band into a rotational
structure of at least ten different branches completely
identified (\cite{Jackson1980}).

These first observations and the models developed to explain
the structure and isophotes of the Ly$\alpha$ emission brought  to the   conclusion that 
the bulk of the production rate of hydrogen and  OH comes from the photodissociation of H$_2$O, which could  be the major 
volatile constituent of comets.

The UV image of comet C/2009 P1 (Garradd) in the OH band was taken in April 1, 2012, when the comet was 229 million kilometers away (\cite{Bodewits2012}). The imagery was made by the Ultraviolet and Optical Telescope (UVOT). UVOT has a 30 cm aperture that provides a $17'\times17'$  field of view with a spatial resolution of 0.5'/pixel in the optical/UV band. Seven broadband filters allow color discrimination, and two grisms provide low-resolution spectroscopy at UV (170 -- 520\,nm) and optical (290 -- 650nm) wavelengths. These grisms provide a resolving power  $\sim100$ for point sources. UVOT   and X-Ray Telescope were used to provide the first ever simultaneous X-ray and UV image of a comet Lulin (\cite{Carter2012}). The UV and X-ray emission are on opposite sides because comet Lulin has two oppositely directed tails.
\cite{Lim2014}  presented observation of both spectra and images of comet C/2001\,Q4 (NEAT) created from the observations  with 
 Far-Ultraviolet Imaging Spectrograph (FIMS), onboard the Korean satellite STSAT-1, (launched  in 2003). FIMS is a dual-channel FUV imaging spectrograph (S-channel 90 -- 115 nm, L-channel
135 -- 171nm , and resolving power $\sim$550 for both channels) with large imaged fields of view (S-channel $4^{\circ}\times$4.6', L-channel $7.5^{\circ}\times$4.3', and angular resolution 5'-- 10') optimized for the observation of the FUV radiation produced by  hot gases in our Galaxy.

The Galaxy Evolution Explorer (GALEX) has observed 6 comets since 2005 (C/2004\,Q2 (Machholz), 9P/Tempel\,1, 73P/Schwassmann-Wachmann\,3 Fragments B and C, 8P/Tuttle and C/2007\,N3 (Lulin)). GALEX is a NASA Small Explorer (SMEX) mission designed to map the history of star formation in the Universe. GALEX's telescopes aperture is 50\,cm. Field cameras were equipped with grism providing  the FUV (134 -- 179\,nm) channel and the NUV (175 -- 310\,nm) channels at resolution 1 -- 2\,nm.  It is also well suited to cometary coma studies because of its high sensitivity and large field of view (1.2 degrees).  \cite{Morgenthaler2009} reported that OH and CS in the NUV were clearly detected in all of the comet data. The FUV  channel recorded data during three of the comet observations and detected the bright CI\,156.1 and 165.7\,nm multiplets. There is an evidence of SI\,147.5\,nm emission in the FUV.
The GALEX data were recorded with photon counting detectors, so it has been possible to reconstruct direct-mode and objective grism images in the reference frame of the comet.

\subsection{Mid and high resolution spectroscopy}

Higher resolution spectroscopy  brings most detailed information on astronomical objects. But it requires rather large apertures and spectrographs with stable structures.  We briefly describe  some findings and technical details of the  larger UV observatories.  

\paragraph{UV observations of comets with the International Ultraviolet Explorer (IUE).}

The operation of the IUE satellite in 1978 -- 1996 was a
real breakthrough in UV observations, in general, and in
comets, in particular. The IUE was able to observe
objects no closer than 45$^{\circ}$ from the Sun, i.e., the comets
that approached the Sun to about 0.7\,AU were accessible.
The stabilization system of the satellite allowed the observation
of moving objects (comets) with an accuracy of
several arcseconds for a 30\,min exposure.  UV spectrographs of
the IUE satellite were practically the only option for
systematic research and monitoring of the spatial distribution
of some of the most common elements in cometary
comas, which are the key to understanding the production
mechanism of these elements. Practically all
comets that reached the brightness of stellar magnitude
6 were observed with the IUE satellite. In
total, 55 comets were systematically studied.

Very briefly, the results of 18.5\,years of UV studies can be formulated
as follows (\cite{Festou1998}):  comets are composed of dust, water,
CO, and CO$_2$. The content of other molecules is a few  per cent of the water content (from
0.5\% to 1\% in comet Hale-Bopp). Chemical composition
of periodic and aperiodic comets differs by no
more than a factor of two and also differs by the gas-to-
dust mass ratio (see however more recent data presented in Fig.\ref{fig1}).  New species were  discovered in the coma, and some of 
the most abundant nuclear species have been studied through the spatial distribution of their dissociation products. 
Time variations from few moths to hours were  investigated in detail.

\paragraph{UV studies of comets with the The Hubble Space Telescope (HST).}

HST is currently
the only  option for obtaining UV cometary
spectra in the 115 -- 300\,nm range. 
44  comets were observed with the HST (till 2016).

The  Space Telescope Imaging Spectrograph (STIS)
which allows registration of spectra in a long-slit
mode is a powerful tool for studying the gas in the
inner parts of the coma with a high spatial resolution.
This area is essential for cometary research, as the gas
in these parts of comets might not be in the radiative
equilibrium state; additionally, complex and little
studied chemical reactions can take place there. Like the IUE the HST has a significant limitation for cometary research. 
This is the Sun avoidance angle which does not allow observations of objects closer than $50\circ$ from the
Sun, i.e., approximately 0.8\,AU. 

\cite{Feldman2016} summarised the results obtained with  HST for four  comets  observed in the FUV with the Cosmic Origins Spectrograph (COS): 103P/Hartley\,2, C/2009\,P1 (Garradd), C/2012\,S1  (ISON) and C/2014\,Q2 (Lovejoy). The principal objective was to determine the relative CO abundance from measurements of the CO Fourth Positive system in the spectral range of 140 -- 170\,nm. In the two brightest comets, nineteen bands of this system were clearly identified. The water production rate was derived from observations of the OH (0,0) band at 308.5\,nm by the STIS. The derived CO/H$_2$O production rate ratio ranged from ~0.3\% for comet Hartley to ~20\% for comet Garradd. 

Still most enigmatic cometary objects studied with the  HST in last decade are the main belt comets (MBC). MBCs  exhibit comet-like mass loss resulting from
the sublimation of volatile ice even though they occupy orbits in the main asteroid belt. Till now there is no definite understanding on  how these cometary bodies appeared in the main asteroid belt.

Of course field cameras of the HST were used for the observation of comets though the small field of view was a serious limitation.

\paragraph{Comets with Far Ultraviolet Spectroscopic Explorer (FUSE).} 
The FUV  region shortward of 200\,nm, contains the resonance
transitions of the cosmically abundant elements, as well as the electronic
transitions of the most abundant  simple molecules such as CO and H$_2$. The
principal excitation mechanism in the ultraviolet is resonance
fluorescence of solar radiation.

FUSE, launched in 1999 June, provided an orbiting capability for the temperature and density diagnostics of molecular species. 
FUSE spectral resolution was  better than 0.04\,nm
in the wavelength range of 90 -- 118.7\,nm 
together with very high sensitivity to weak
emissions, making possible both the search for minor coma species and
extensive temperature and density diagnostics for the dominant
species. \cite{Feldman2001,Feldman2002}
reported observations of comet C/2001\,A2 (LINEAR) with FUSE. Several new cometary emissions were identified. 

\section{Future observations of comets with World Space Observatory -- Ultraviolet (WSO-UV).} 

Observations of comets from space have several specific features. As per \cite{Sachkov2016b} these are:

\begin{itemize}

\item Comets are moving objects. Their velocities can
reach tens of arcseconds per hour. For example, comet
Halley moved at a  velocity of 11''/hour, relative
to the stars. For obtaining the UV spectra, it is necessary
 to support the spacecraft stabilization
during long exposures, having a priori knowledge of this velocity.

\item The presence of solar lines in
the spectrum significantly complicates the observation
for some cometary velocities since both 
solar and cometary spectra, are recorded together.
The observed solar spectrum is shifted by the
comet's heliocentric velocity, the so-called Swings effect.

\item Comets are variable objects.
Cometary nuclei produce short-lived tails and comas. Monitoring of this temporal changes 
requires  continuous or quasi-continuous observations.

\item Comets are extended objects. Larger fields of view of cameras onboard space observatory are preferable. 
For (imaging) spectrography long-slit spectrographs are the most suitable. 

\item Comets are faint objects. Therefore, high resolution spectroscopic  observations require for  sufficiently large aperture telescopes.
 
\item An important restriction in cometary studies from the near-Earth orbit is the significant sun avoidance angle 
which limits the observations during perihelium passage.

\end{itemize}

The World Space Observatory -- Ultraviolet (WSO-UV) fits  most of the  requirements listed above. WSO-UV is a multi-purpose international space mission born as a response to the growing up demand for UV facilities by the astronomical community. WSO-UV was described in previous publications in great detail  (see \cite{Shustov2011, Shustov2014, Sachkov2014,  Sachkov2016}, for this reason only  basic information on the project, relevant to the observation of comets,  is  briefly presented here. 

The main scientific purpose of WSO-UV is the spectroscopic observation of faint UV sources and high resolution UV imaging.  The parameters of the T-170M telescope
(large 170 cm diameter primary mirror, UV optimized coating of the optical surfaces,
high-accuracy guidance and stabilization system, etc.)
were chosen to fit requirements of high angular resolution and maximum effective area in the 110\,nm -- 320\,nm
range. 
Construction of the telescope  provides  the solar avoidance angle about $40^{\circ}$. This is important for observations of comets at low angular distance from the Sun.

\begin{itemize}
\item high-resolution spectroscopic
observations of point-like objects in the 110 -- 320\,nm spectral
range. Resolving power of  the two high resolution spectrographs  designed for observations in  the NUV
 (180 -- 320\,nm) and FUV (110 -- 180\,nm) is   $R > 50000$;
\item low-resolution (resolving power $R\sim1000$) spectral
observations. A long slit spectrograph   will operate in the 110 -- 320\,nm range too. It is especially suitable for  faint and extended objects such as comets;
\item solar-blind FUV imaging with angular  resolution (0.08'') for direct imaging with capabilities for low dispersion slit less spectroscopy; 
\item wide field NUV-visible imaging with field of view 10'$\times$7.5' and angular resolution 0.15''; 
\end{itemize}

Broadly speaking,  the spectroscopic capabilities of WSO-UV and HST are rather similar; in the NUV range, WSO-UV spectrographs will be more efficient than the HST/STIS spectrograph  however, the sensitivity of HST/COS in the FUV range will not be matched by WSO-UV. To evaluate the impact of the design on the scientific objectives of the mission, a simulation software tool has been developed for the WUVS (\cite{MarcosArenal2017}). This simulator builds on the development made for the PLATO space mission and it is designed to generate synthetic time-series of images by including models of all important noise sources.

\begin{figure*} [ht!]
	\centering
	\includegraphics[width=0.8\linewidth]{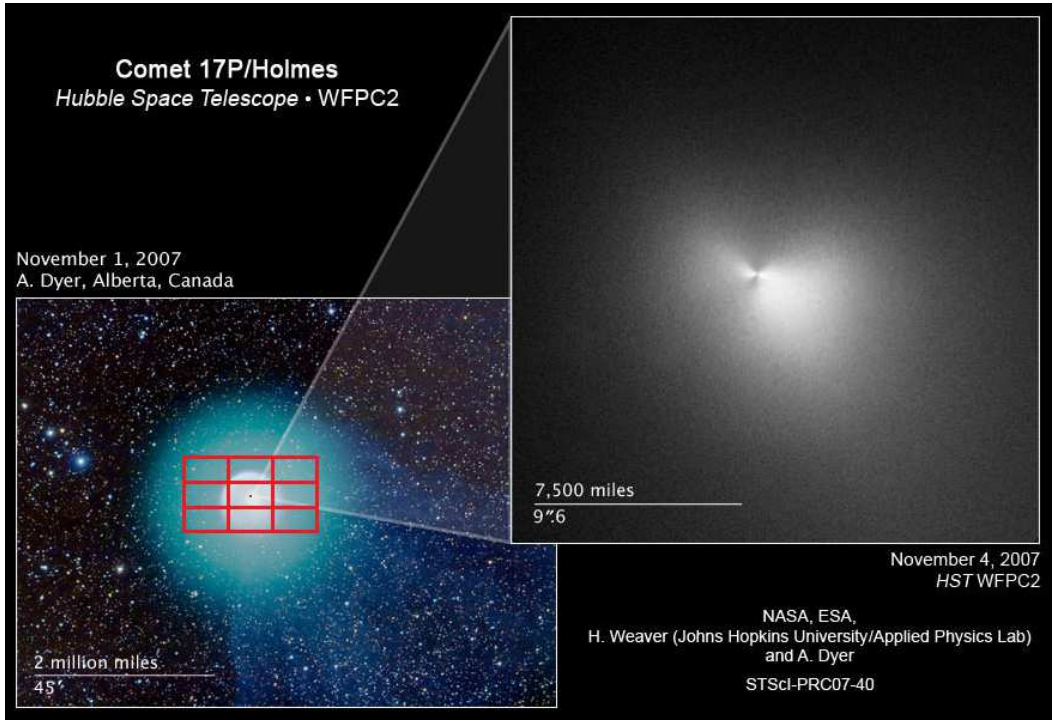}
	\caption{
	Comparison between the field of view (FoV) of the  WSO-UV NUV imager and the FoV of the planetary camera PC1 WFC-2 on HST. The Hubble image and field of comet 17P/Holmeshas been taken from https://apod.nasa.gov/apod/ap071128.html.  Each red rectangle in the lower left inset represents the FoV of the WSO-UV NUV imager. Notice that just 9 exposures are require to cover the inner coma at full.
	} 	
\label{fig3}
\end{figure*}

Some  characteristics of the Field Camera Unit (FCU) channels significantly differ from those of  HST as shown  in Tabl.\ref{tab1} (Hubble data are taken from HST Instrument Handbooks).	
In Fig.\ref{fig3} we illustrate the difference between the field of view of the NUV imager  and the high resolution camera PC1 WFPC-2 on board  HST.

 \begin{table*}[ht!]
 	\centering
 	\small
 	\caption{Comparative characteristics of	the	 NUV and FUV channels of the FCU with those of   HST/ACS/SBC  and HST/WFC3/UVIS} 
 	\label{tab1}
 	\begin{tabular}{@{}lrrrr@{}}
 		\tableline
 		Parameters & FUV Channel & NUV Channel & HST/ACS/SBC & HST/WFC3/UVIS \\
 		\tableline
 		Detector & MCP& CCD & MCP,	MAMA & CCD\\
 		Spectral	range,	nm & 115 \sbond 170 & 174\sbond 310	(ext. to 1000) &115\sbond 170 & 200\sbond 1000 \\
 		D	(primary	mirror),	m &1.7 &1.7  &2.4  &2.4 \\
 		Field	of	view, arcsec & $163\times 163$ &$597\times 451$	  &$35\times 31$  &$162\times 162$\\
 		Scale,	arcsec/pixel  & 0.08$^{[a]}$  & 0.146  &$0.033\times 0.030$	 & 0.0395 \\
 		Detector	size,	mm & 30	 & $49\times$ 37 &25	 & $61\times$ 61 \\
 		Detector	format & 2k$\times$2k & 4k$\times$3k	 &1k$\times$1k & 1k$\times$1k \\
 		Number	of	filters & up	to 10+2 prims	 & upto 15 &6+2	prism & 62 \\
 		\tableline
	[a]  Angular resolution. & & & & \\
 	\end{tabular}
 \end{table*}

Similar to \cite{Weaver2002} we 
consider that the main scientific objective is to obtain accurate abundance measurements for all known UV-emitting cometary species CO from the CO\,4PG bands, C2 from the CO Cameron bands, S2 from the S2 B-X bands, CS2 from CS emissions, and water from OH emissions and to perform a deep search for any previously undetected species. The long slit capability of the  WSO-UV will allow us to characterize the spatial distribution of the coma species, so that we can identify those derived from an extended source e.g. CO, study the decay of short-lived species e.g. S2, and investigate the importance of electron impact on CO for the excitation of the Cameron bands. The latter issue can be definitively resolved with high spectral resolution observations of any comet having V$<$5. If an exceptionally bright (V$<$2) comet is discovered, we would then request Director's  time to measure the D/H ratio. The D/H ratio in cometary water is a key indicator of the role played by comets in the delivery of volatiles to the terrestrial planets. The deuterium abundance of water in comets preserves information about the formation conditions of our solar system, while also constraining the possible contribution of cometary water to Earth's oceans. The D/H values in Jupiter Family (JF) comets and the ice ratios (CO/CO2/H2O) in both JF  and Oort Cloud comets call into question the details of the dynamical processes for populating the modern reservoirs for these two dynamical classes. \cite{Bodewits2015} noted that currently too few D/H ratios have been measured in comets to allow for a meaningful interpretation.

Contemporary interpretation of the term ``comets'' includes so called exocomets.  We remind  that exocomets (cometary activity in extrasolar systems) have been first detected in optical observations by the
temporal  variations of the CA\,II-K line (\cite{Ferlet1987}) and confirmed in UV observations spectra of the well-studied $\beta$-Pictoris debris disk with the HST  (\cite{Vidal-Madjar1994}). In recent years many interesting results were obtained by intensive optical monitoring of high-resolution optical spectra of the Ca II lines in gaseous disks around selected stars.  We think  that UV monitoring can be also  valuable for future  studies of exocomets.  \cite{Miles2016} presented  the analysis of time-variable Doppler-shifted absorption features in FUV spectra of the unusual 49 Ceti debris disk. This nearly edge-on disk is one of the brightest known and is one of the very few containing detectable amounts of circumstellar (CS) gas as well as dust.   \cite{Miles2016}  calculated the velocity ranges and apparent column densities of the 49 Cet variable gas, which appears to have been moving at velocities of tens to hundreds of km/s relative to the central star. The velocities in the redshifted variable event  showed that the maximum distances of the infalling gas at the time of transit were about 0.05 -- 0.2\,AU from the central star. A preliminary composition analysis brought to conclusion that the C/O ratio in the infalling gas is super-solar, as it is in the bulk of the stable disk gas. 

Measuring the evolution of gas and small bodies in the young planetary disks will help us understanding the 
formation of structures such as the Oort cloud and the Kuiper belt. However, this requires extensive monitoring programs that are very inefficient when carried out from low Earth orbit observatories like Hubble.
WSO-UV should be able to run several monitoring programs simultaneously in a cost-efficient manner.

Comets aggregate from interplanetary dust and there are growing evidences that the large organic molecules found in the cometary samples come from the interstellar dust trapped in the pre-solar nebula (see e.g. \cite{Bertaux2017}).   The UV spectrum  is  very sensitive to small particles and large molecules.   WSO-UV imagers and filters sets are designed to measure variations of the extinction curve in the coma and together with the long slit spectrograph will allow characterizing variations between the comets populations in particle size distribution and molecular bands. WSO-UV FUV imager will be operated from high Earth orbit facilitating the monitoring and minimizing the contamination from the Earth geocorona. 

\section{Conclusions}

Comets are   important ``eyewitnesses'' of  the processes involved in  Solar System formation and evolution.  Observations of the UV spectra of comets from rockets and satellites have brought to light some new clues on those primitive bodies that could be the link between interstellar molecules and the early planetary atmospheres,
as well as the bridge between stellar and planetary astrophysics.  Some ill known physical processes require to be addressed the acquisition of the data only attainable in the UV range of the electromagnetic spectrum. Thus, space-born facilities are required to study comets; complementary ground-based observations and in-situ experiments
are also needed to get a comprehensive view of the problem. 

We briefly reviewed half a century of UV observations of comets and argue that it  was a very fruitful period for understanding the complex nature of comets  and their role in the evolution of the Solar System. Both short- and long-term experiments made an invaluable contribution to the development of cometary science. Although research on comets in situ (such as the remarkable Rosetta �project) yield direct and very important data, the remote study of comets will retain its significance for many decades. 

The WSO-UV mission will guarantee the continuity of the UV observation of comets, as well as exocomets. The large FoV and high sensitivity of the NUV imager together with the high Earth orbit will made of WSO-UV the most efficient observatory ever flown to track comets evolution.

\acknowledgments
B.Shustov,  V.Dorofeeva and E.Kanev thank the Russian Science Foundation for supporting this work by grant RSF 17-12-01441.
Ana I G\'omez de Castro and Juan C. Vallejo thank the Spanish Ministry of Economy, Industry and Competitiveness for grants:   ESP2014-54243-R, ESP2015-68908-R.

\nocite{*}
\bibliographystyle{spr-mp-nameyear-cnd}
\bibliography{COMETS}

\end{document}